\documentstyle[aps,epsf]{revtex}

\input epsf

\newcommand{\half}{{1\over2}}
\newcommand{\be}{\begin{equation}}
\newcommand{\ee}{\end{equation}}
\newcommand{\beq}{\begin{equation}}
\newcommand{\eeq}{\end{equation}}
\newcommand{\bea}{\begin{eqnarray}}
\newcommand{\eea}{\end{eqnarray}}
\newcommand{\beal}{\setcounter{letter}{1} \begin{eqnarray}}
\newcommand{\eeal}{\addtocounter{equation}{1} \end{eqnarray}}
\newcommand{\none}{\nonumber \\}

\newcommand{\gbar}{\overline{g}}
\newcommand{\phibar}{\overline{\phi}}

\newcommand{\p}{\partial}

\begin{document}
\draft

\title{\bf Spherically symmetric scalar field collapse in any
dimension}

\author{M. Birukou${}^\sharp$, V. Husain${}^\flat$, G.
Kunstatter$^\sharp$,
 E. Vaz${}^\sharp$, and M. Olivier$^\dagger$}

\address{$\flat$Dept. of Mathematics and Statistics\\
University of New Brunswick,
Fredericton, N.B. Canada E3B 1S5.\\
$\sharp$Dept. of Physics and Winnipeg Institute of
Theoretical Physics\\
University of Winnipeg,
Winnipeg, Manitoba Canada R3B 2E9.\\
$\dagger$Dept. de Physique, \\Universite Laval,
Quebec, QC, Canada G1K 7P4}

\date{\today}
\maketitle

\begin{abstract}

We describe a formalism and numerical approach for studying spherically
symmetric scalar field collapse for arbitrary spacetime dimension $d$
and cosmological constant $\Lambda$. The presciption uses a double null
formalism, and is based on field redefinitions first used to simplify
the field equations in generic $2-$dimensional dilaton gravity. The
formalism is used to construct code in which $d$ and  $\Lambda$ are
input parameters. We reproduce known results in $d=4$ and $d=6$  with
$\Lambda = 0$, and present new results for $d=5$ with zero and negative
$\Lambda$.

\end{abstract}


\section{Introduction}\medskip

It is an interesting fact that spherically symmetric gravitational
collapse exhibits critical behaviour \cite{choptuik}. This is a
classical effect in phase space: there are one parameter families of initial 
data sets, for a variety of matter fields, such that as the parameter is
tuned, a transition from reflection of infalling matter to black hole formation
is observed numerically. Two types of behaviour are observed close to
this transition point. Depending on the matter type, black holes form
with zero or non-zero initial mass, and the matter field exhibits
discrete or continuous self-similarity. A variety of matter fields have been 
studied since the initial seminal work by Choptuik. There have also been some
extensions beyond spherical symmetry, as well as a semi-analytic perturbation
theory understanding of the critical exponent. (Recent reviews may be found in
Refs.\cite{gund,lehn}.) If an exact time dependent solution were
available with the appropriate boundary conditions, this critical behaviour would
be manifested in the solution of the apparent horizon condition,
which is a transcendental equation \cite{hmn1}. However, to date only one
time dependent (and non-self similar solution) is known, but this does not 
have the required asymptotically flat boundary conditions \cite{hmn2}. Thus
the full PDE problem must be studied numerically.

Most numerical studies of spherically symmetric collapse have been
in four spacetime dimensions. The only exceptions are the massless
minimally coupled scalar field by Garfinkle\cite{garf_sixd} in six
spacetime dimensions, and by Pretorius and Choptuik \cite{pc}, and
Husain and Olivier\cite{husain} in three dimensions with negative
cosmological constant. In addition, there exist two papers that
study this problem in the much simpler case of the self-similarity
ansatz, an ODE problem, for any spacetime dimension
\cite{soda,frolov}.

The purpose of this work is to present a formalism and numerical
method for studying the gravitational collapse of a spherically
symmetric scalar field for any value of spacetime dimension and
cosmological constant. The approach reproduces and supplements known 
results in four and six dimensions with zero cosmological constant, 
and gives new results in five spacetime dimensions with zero and 
negative cosmological constant.

\section{Field Equations}
\par
The basic idea for obtaining field equations valid for any dimension is
to reduce the $d-$dimensional Einstein-scalar field equations by
imposing spherical symmetry, and then use a field redefinition originally 
motivated by generic 2-dimensional dilaton gravity \cite{dil_grav}. This 
allows the  Einstein equations for $d-$dimensional, spherically symmetric 
scalar field collapse to be put into a form that can be managed numerically 
by a single code. In actuality, the formalism and code we describe below is 
applicable to a more general class of models that includes non-minimal 
scalar field coupling. This class of theories is quite broad, and 
contains as a sub-class the $d-$dimensional spherically symmetric case.

Einstein gravity with cosmological constant in $d$ spacetime 
dimensions is given by the action
\beq
S^{(d)}_G={1\over 16\pi G^{(d)}}\int d^dx\sqrt{-g^{(d)}}
   (R(g^{(d)}) - \Lambda)
\label{Einstein}
\eeq
The corresponding action for a minimally coupled scalar field is
\beq
S_M=-\int d^dx\sqrt{-g^{(d)}}\left(
g^{(d)\mu\nu}\partial_\mu\chi\partial_\nu
\chi\right)
\label{matter}
\eeq
To impose spherical symmetry, we write the $d-$dimensional metric
$g_{\mu\nu}$ as
\be
ds^2_{(d)} = \bar{g}_{\alpha\beta} dx^\alpha dx^{\beta} + r^2(x^\alpha)
d\Omega_{(d-2)},
\ee
where $d\Omega_{(d-2)}$ is the metric on $S^{d-2}$ and $\alpha,\beta =
1,2$.
This gives the reduced action
\footnote{In most cases one should perform dimensional reduction at the
level of the field equations to guarantee that one obtains the correct
solution space. It is well known\cite{dil_grav} that in the present case
the reduced field equations correspond to the field equations obtained
from the reduced action.}
\bea
S_{TOT}&:=& S_G + S_M \none
&=& {{\cal V}^{(n)}} \int d^2x \sqrt{-\gbar}\left[{1\over 2G}
\left({n\over 8(n-1)} \phibar^2(R(\gbar)
-\Lambda)+{1\over 2}|\partial\phibar|^2 + {n^2\over 8l^2}
\phibar^{2n-4\over n} \right)\right.\none
 & &\left. - l^n \phibar^2
| \partial\chi|^2\
\right]
\label{reduced1}
\eea
where $n=d-2$, ${\cal V}^{(n)}$ is the volume of the unit $n$-sphere,
and
\bea
l^n= G^{(d)}\\
{1\over 2G}= {8(n-1)\over 16\pi n}\\
\phibar = \left({r\over l}\right)^{n\over2}
\eea

The key simplification in our formalism is achieved by the following
conformal reparametrization of the metric, which eliminates the kinetic
term for $\bar{\phi}$ from the action \cite{dil_grav}. Let
\beq
g_{\alpha\beta} = \Omega^2(\phibar) \gbar_{\alpha\beta},
\label{metric}
\eeq
where
\beq
\Omega^2(\phibar) = C\exp\left(\half \int {d\phibar\over
(d\overline{D}/d\phibar)}\right),
\label{Omega}
\eeq
\beq
 \overline{D}(\phibar) = {n\over 8(n-1)}\phibar^2
\eeq
and $C$ is an arbitrary constant. Now define a dimensionless
``dilaton'' field
\beq
\phi = \overline{D}(\phibar) = {n\over 8(n-1)}\left({r\over l}\right)^n
\label{dilaton}
\eeq
Note that $\phi$ is proportional to the area of the $n$-sphere at
radius $r$. With these redefintions the reduced action takes
the simpler form
\bea
S_{TOT}= {1\over 2G}\int d^2x \sqrt{-g}\left[\phi R(g) +
V^{(n)}(\phi)\right]
-\int d^2x\sqrt{-g}\ H^{(n)}(\phi) |\partial \chi|^2
\label{dilatonaction}
\eea
where
\bea
H^{(n)}(\phi)&\equiv& {8(n-1)\over n} \phi\\
V^{(n)}(\phi)&\equiv& {1\over n} \left({8(n-1)\over n} \right)^{1\over
n} \phi^{1\over n}
\left(-l^2 \Lambda
+{n^2\over 8}\left({8(n-1)\over n}\right)^{n-2\over n}
   \phi^{-2/n}\right),
\eea
and the overall factor of ${\cal V}^{(n)}$ has been dropped.

For arbitrary functions $V(\phi)$ and $H(\phi)$,  the
action (\ref{dilatonaction}) is that of generic dilaton gravity
theory coupled to a scalar field in two spacetime
dimensions. This theory has been studied in great
detail \cite{dil_grav}. The vacuum equations ($\chi=0$) can be solved
exactly. By choosing an adapted coordinate system in which the the
dilaton $\phi$ plays the role of the spatial coordinate (i.e.
$x=l\phi$), the vacuum solution for the metric is
\beq
ds^2 = -(j(\phi)-2GM)dt^2 +(j(\phi)-2GM)^{-1}dx^2
\label{solution}
\eeq
where $M$ plays the role of  mass, and
\beq
j(\phi)\equiv \int^\phi_0 d\tilde{\phi} V(\tilde{\phi})
\eeq
Note that we have dropped the superscript $n$ denoting spacetime
dimension, since the above solution applies to the generic case.
For the specific case of $d-$dimensional spherically symmetric gravity
\bea
j^{(n)}(\phi) &=& {1\over n}\left({8(n-1)\over n} \right)^{1\over n}
 \times \nonumber \\
 &&  \left(-l^2 \Lambda \left({n\over n+1}\right)\phi^{n+1\over n}
 + {n^3\over 8(n-1)}\left({8(n-1)\over n}\right)^{n-2\over n}
   \phi^{n-1\over n}\right)
\eea
It can easily be verified by making the appropriate substitutions and
conformal reparametrization (\ref{metric}) that the
physical line element $ds^2_{(d)}$ corresponding to (\ref{solution})
is precisely that of a $d$ dimensional deSitter/anti-deSitter black hole
with mass $M$.\footnote{In order to get the overall scale factor right, 
one must choose the  constant $C$ in (\ref{Omega}) appropriately.}
It is important to note that the metric (\ref{solution}) is singular at
$\phi=0$ even when $M=0$. Up to numerical constants, $j$ goes to zero as
\beq
j(\phi)\to \phi^{1-1/n}
\eeq
near $\phi=0$. This is not a physical singularity since the
physical metric $\overline{g}$ is indeed the Minkowski metric when $M$
and
$\Lambda$ are zero. Nonetheless, the vanishing of $j(\phi)$ will affect
the choice of boundary conditions in our numerical method.

We now examine the field equations that derive from
(\ref{dilatonaction}) in double null coordinates, for which the metric
may be
parametrized as
\beq
ds^2 = - 2 l g(u,v)\phi'(u,v) du dv
\label{double null}
\eeq
where the prime denotes partial differentiation with respect to the null
coordinate $v$. (Recall that this is just the $u-v$ part of the physical
metric.) The corresponding field equations are
\bea
& &\dot{\phi}' = - {l\over 2} V^{(n)}(\phi) g \phi'
\label{double null equations a}\\
& &{g'\phi'\over g H^{(n)}(\phi)} = 2G (\chi')^2 \label{cons}\\
& &(H^{(n)}(\phi)\chi')^{\cdot} + (H^{(n)}(\phi) \dot{\chi})' = 0.
\label{double null equations c}
\eea
In the above, the dot refers to differentiation with respect to $u$,
which is treated like the ``time'' coordinate for the purposes of the 
following numerical integration. Remarkably, for arbitrary $n$
Eqs. (\ref{double null equations a}-\ref{double null equations c}) are
virtually identical in form to those studied in \cite{husain} in
the context of 2+1 dimensional AdS gravity.\footnote{For $n=1$ they are
identical as expected.} However, the boundary conditions are
special in the case of $n=1$ ($d=3$), so we will not consider this case
further. Except where explicitly stated, we henceforth restrict
consideration to $n\geq 2$ (which means spacetime dimension $d \ge 4$).

The evolution equations may be put in a form more useful for numerical
solution by defining the variable
\be h = \chi + {2\phi\chi' \over \phi'}.
\label{def h}
\ee
This effectively replaces the scalar field $\chi$ by $h$.
The evolution equations are
\bea
\dot{\phi} &=& -\tilde{g}/2 \label{phidot}\\
\dot{h} &=& {1\over 2\phi} (h - \chi)\left( g\phi V - \tilde{g}\right),
\label{hdot}
\eea
where
\be
\tilde{g} = \int _u^v (g \phi'V) dv'.
\label{gbar}
\ee
and $\chi$ is now to be considered a functional of $h$ and $g$ given by
\bea
\chi={1\over
2\sqrt{\phi}}\int^v_udv\left[{h\phi'\over\sqrt{\phi}}\right]
+{K_3(u)\over \sqrt{\phi}}
\label{chi integral}
\eea
The integration constant $K_3(u)$ must be zero because the
definition of $h$ requires $h=\chi$ at $\phi=0$.\footnote{We will see
below that although $\phi'$ goes to zero at $\phi=0$ it does so slowly
enough to guarantee that the second term in (\ref{def h}) vanishes 
at $\phi=0$.} The function $g$ is a functional of $h$ and $\phi$, 
obtained by integrating the constraint (\ref{cons}):
\beq
g=K_1(u)\exp\left[4\pi\int^v_udv{\phi'\over\phi}(h-\chi)^2\right]
\eeq
where $K_1(u)$ is again an integration constant (i.e. independent of 
the ``spatial'' coordinate $v$).
We consider the case of a spherically symmetric, collapsing
shell of matter, with no black hole in the interior, initially. Thus,
our boundary condition should be such as to guarantee that
the metric at $r=0$ (which translates to $\phi=0$) goes over to the
vacuum solution (\ref{solution}). By transforming the vacuum ($M=0$)
metric
(\ref{solution}) to double null coordinates (\ref{double null}) we
obtain a
metric of the form
\beq
ds^2 = -j(\phi)dudv
\label{metric 3}
\eeq
where $u=t-\phi^*$ and $v=t+\phi^*$, with the generalized
``tortoise coordinate'' $\phi^*$ defined by
\beq
\phi^*=l\int^\phi_0{d\phi\over j(\phi)}
\label{phistar}
\eeq
With these definitions, $\phi=0$ corresponds to the surface
$v=u$. Moreover, it follows that for the vacuum solution
\beq
\phi'\equiv{\partial \phi\over \partial v} = {1\over 2}{j(\phi)\over l}
\eeq
Comparing metric (\ref{metric 3}) to our general form (\ref{double
null}) we see that $g=1$ for the vacuum solution. Since we would like
the numerical solution to approach the vacuum at $\phi=0$, the above
analysis determines the required boundary conditions. In particular the
integration constant $K_1(u)=1$, and
\beq
\dot{\phi}\equiv{\partial \phi\over \partial u}\to  -{1\over
2}{j(\phi)\over l}
\eeq
which vanishes at $\phi=0$ in agreement with the expression
(\ref{gbar}).

\section{Numerical Method}
The numerical scheme uses a $v$ (`space') discretization to obtain a
set of coupled ODEs:
\be
h(u,v) \rightarrow h_i(u),\ \ \ \ \ \ \ \phi(u,v)\rightarrow \phi_i(u).
\ee
where $i = 0,\cdots, N$ specifies the $v$ grid. Initial data for these
two
functions
is prescribed on a constant $v$ slice, from which the functions
$g(u,v),\tilde{g}(u,v)$ are
constructed. Evolution in the `time' variable $u$
is performed using the 4th. order Runge-Kutta method. The general 
scheme is similar to that used in \cite{GP}, together with some 
refinements used in \cite{DG}. This procedure was also used for the 
$3-$dimensional collapse calculations in \cite{husain}.

The initial scalar field configuration $\chi(\phi,u=0)$ is most
conveniently specified as a function of $\phi$ rather than $r$. (Recall 
that $\phi\propto r^n$.)  This together with the initial arrangement of 
the radial points $\phi(v,u=0)$ fixes all other functions. We used the initial
specification $\phi(0,v) = v$.

We consider two types of initial scalar field configurations: the
Gaussian
and ``tanh'' functions
\be
\chi_G(u=0,\phi) = a \phi\ {\rm exp}\left[-\left(\ {\phi-\phi_0\over
\sigma}\right)^2\right],
\ee
and
 \be
\chi_T(u=0,\phi) = a \tanh(\phi).
\ee
These choices permit us to test ``universality,'' which is defined to be
the independence of the details of the collapse, such as the critical
exponent, on the choice of initial data shapes and parameters. Although
universality may be tested for a large variety of shapes and parameters,
the emphasis here is on different spacetime dimensions, so we have
restricted attention to varying the amplitude parameter for these two shapes.

The initial values of the other functions are
determined in terms of the above by computing the
integrals for $g_n$ and $\tilde{g}_n$ using Simpson's rule.
In all cases, we used values of $\phi_0=1$ and $\sigma =0.3$ for the
Gaussian initial data.

The boundary conditions at fixed $u$ are
\be
\phi_k=0, \ \ \ \tilde{g}_k=0,\ \ \ \ g_k=1.
\ee
{\it where $k$ is the index corresponding to the position of the origin
$\phi=0$.} (In the algorithm used, all grid points $ 0\le i \le k-1$
correspond to ingoing rays that have reached the origin and
are dropped from the grid; see below). These conditions are equivalent
to $r(u,u)=0$, $g|_{r=0} = g(u,u) =1$, and guarantee regularity of
the metric at $r=0$.  Notice that for our initial data, $\phi_k$ and
hence $h_k$ are initially zero, and therefore remain zero at the origin
because of Eqn. (\ref{hdot}).

As evolution proceeds via the Runge-Kutta procedure, the entries in the
$\phi_i$ array sequentially reach 0, at which point they are
dropped from the grid. Thus the radial grid loses points with evolution.
This is similar to the procedure used in \cite{GP} and \cite{DG}.

At each $u$ step, a check is made to see if an apparent horizon has
formed by observing the function
\be
ah\equiv {g}^{\alpha\beta}\p_\alpha \phi \p_\beta \phi =
-{\dot{\phi}\over lg},
\label{aheqn}
\ee
whose vanishing signals the formation of an apparent horizon.
\footnote{It is worth noting that there exist special foliations 
of black hole spacetimes which have no apparent
horizon\cite{wald}. Therefore there is in principle the possibility 
that a numerical scheme that encounters such a slicing may
fail to detect black hole formation. This manifestly does not happen
with the double null coordinates used in our study, as demonstrated
by the form of the static solution described above, and our results
below.} For each run of the code with fixed amplitude $a$, this function is
scanned from larger to smaller radial values after each  Runge-Kutta
iteration, and evolution is terminated if the value of this function
reaches $10^{-3}$. The corresponding radial coordinate value is recorded
as $R_{ah}$. In the subcritical case, it is expected that all the radial
grid points reach zero without detection of an apparent horizon. This 
is the signal of pulse reflection.

The results $(a, R_{ah})$ are collated as in \cite{choptuik},
by seeking a relationship of the form
\beq
R_{ah}\propto (a-a_*)^{\gamma}
\label{sl}
\eeq
where $a_*$ is the critical amplitude which separates the black hole 
and reflection solutions. $R_{ah}$ is the basic dimensional scale if
a black hole forms, and is linearly related to black hole mass in four
spacetime dimensions.

To improve numerical accuracy near $\phi=0$ we follow a procedure
similar to \cite{DG}, where all functions on a constant $u$ surface are
expanded in power series in $\phi$ at $\phi=0$, and the first three
values
of the constraint integrals are derived using the respective power
series.
We write
\be
h = h_0 + h_1 \phi
\ee
and calculate the parameters $h_0,h_1$ using the linear least squares
fit for the first fifteen points in $h_i(u)$.\footnote{There is
nothing fundamental about this number, since the behaviour of the scalar
field turns out to be very linear over the first several points; the
results for $h_0$ and $h_1$ were virtually insensitive if the number of
points used varied by a few on either side of fifteen.} From this the
expressions for $\chi$, $g$ and $\tilde{g}$ follow. The remaining $N-3$
values of these functions are computed from their  integral using
Simpson's rule for equally spaced points. This linear fit near the $\phi=0$ 
is necessary because  it elegantly handles the problematic $1/\phi$ factor
in the $h$ evolution equation, which would persist even if a finer mesh
were used. The $v-$derivatives of functions (needed for computing $g$
and
$\tilde{g}$) are calculated using $f'_i = (f_{i+1} - f_{i-1})/{2\Delta
v}$
with end point values determined by linear extrapolation:
$f'_1 = 2f'_2 - f'_3$ and $f'_{N} = 2f'_{N-1} - f'_{N-2}$.

Further comments on the procedure are the following.
The number of $v$ grid points decreases as ingoing null geodesics cross
$r=0$, and so a reflected pulse cannot be followed back out toward
infinity.\footnote{A modification of our procedure along the lines 
suggested in \cite{lehn2} may allow the tracking of the reflected 
pulse to future null infinity.} Also, again due to the loss of grid points, 
and hence resolution, we are not able to observe the detailed behaviour 
of the scalar field very near criticality. An additional numerical adjustment
concerns the enforcement of boundary conditions at the origin: there is
the gravitational tendency for the matter to pile up at the origin as
the collapse proceeds. However
the formalism has the competing implicit condition $h(u,u)=0$ (ie. at
$\phi=0$). This can lead to a shift of this boundary condition under
evolution. It is rectified by adjusting the scalar field function at 
each time step by adding a constant shift at all points. This shift is 
of order $10^{-5}$ or less at each time step, and therefore there is a 
minor loss of accuracy, but a corresponding gain in stability.

The code was tested for grid sizes ranging from 2000 to 6000 points, and
with the $u$ and $v$ step sizes ranging from $10^{-2}$ to $10^{-4}$, for
the two types of initial data used, as well as the vacuum case of
vanishing scalar field. These tests established that the code converges. 
Further tests of the code is the reproduction of the known results in four 
and six dimensions, which also demonstrates the accuracy of the results we
obtain. All the results presented below were for a grid size of 6000 points,
with $u$ and $v$ step sizes of $10^{-3}$.

Finally, we point out that this procedure allows a more accurate
analysis of the supercritical case than the subcritical one, again
because of the loss of grid points as the evolution proceeds. This
is less of an issue in the supercritical case because the number of
points lost depends on the initial pulse amplitude, and our termination
condition is such that typically one quarter of the original number of
points are still present at termination, for the closest approach to
criticality. We could of course monitor the subcritical case up to 
this point as well in an attempt to observe the self-similarity  of the
scalar field, but the very nature of this behaviour requires a way to 
replace lost grid points, as observed in \cite{DG}. For this reason we 
focus on supercritical evolutions.

\section{Results and Conclusions} The code was first tested to
recalculate known results in four and six dimensions. It was then run
for the five dimensional case with zero, positive and negative
cosmological constant. All the calculations were performed for
amplitudes above the threshold for black hole formation, and for 
initial data specified in both Gaussian and tanh forms, parametrized by 
amplitude $a$. The figures below show the scaling law Eqn. (\ref{sl}). 
The squares represent the points $(a,R_{ah})$ and the lines are the least 
squares fit to these points.

The $4-$dimensional results for the Gaussian initial data are
illustrated in Figure 1. The least squares fit gives a
slope of $\gamma = 0.36$, in good agreement $(\sim 4\%)$ with earlier
studies \cite{choptuik,gund,lehn}. The figure also shows the oscillation
about the fit line, again in accord with earlier work.

% FIG 1
\begin{figure}[t]
\epsfxsize=130mm
\epsffile{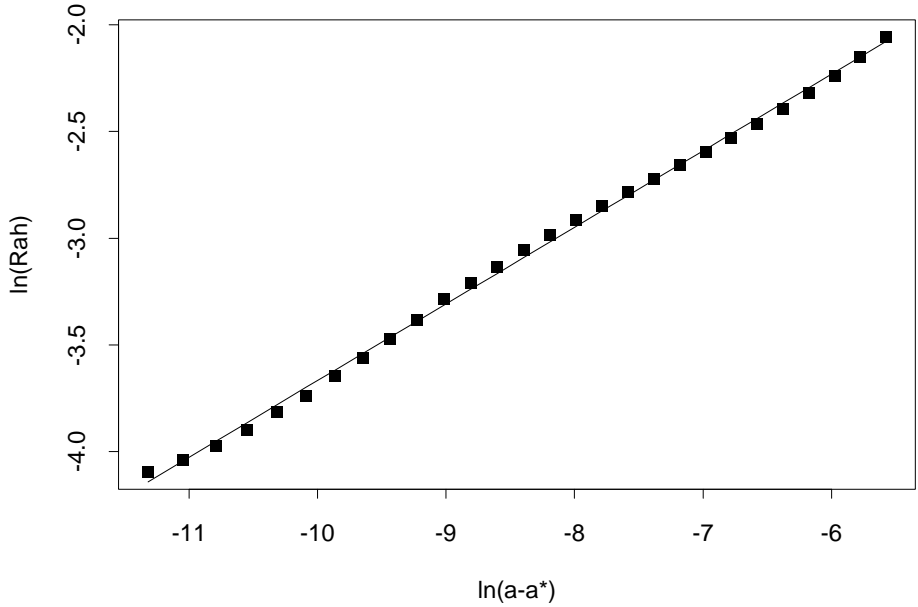}
\caption{\baselineskip = 1.0em Logarithmic plot of  apparent horizon
radius $R_{ah}$ versus initial scalar field amplitude $(a-a_*)$  
in four spacetime dimensions for Gaussian initial data. $\gamma = 0.36$}
\end{figure}

The 6-dimensional results for Gaussian data appear in Figure 2. Our
result for the critical exponent is $\gamma = 0.44$. For comparison,
the result in Ref. \cite{garf_sixd} is $\gamma = 0.424$.

%FIG 2
\begin{figure}[t]
\epsfxsize=130mm
\epsffile{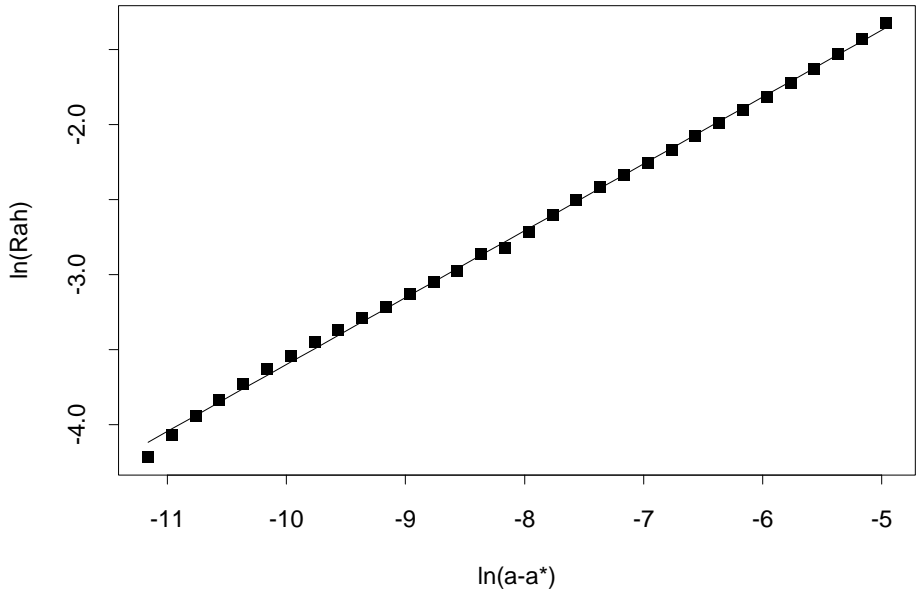}
\caption{\baselineskip = 1.0em Logarithmic plot of apparent horizon
radius $R_{ah}$ versus
initial scalar field amplitude $(a-a_*)$  in six
spacetime dimensions for Gaussian initial data. $\gamma = 0.44$}
\end{figure}

The results for the four and six dimensional calculations for the
tanh initial data are $\gamma = 0.35$ and $\gamma = 0.41$ respectively.
This provides a further check of our code, and further evidence of the
insensitivity of the critical exponent to the shape of the initial data
(``universality'') in both four and six dimensions. Note that in six
dimensions this is the first evidence for universality since
Ref. \cite{garf_sixd} contains results only for a specific gaussian 
form of initial data, different from the one used here. Note also that 
the agreement of our apparent horizon radius scaling results with those 
of the earlier works cited, shows that the apparent horizon appears to 
be a fairly good approximation to the event horizon of the long time 
static limit, (in sofar as these earlier studies actually find this limit).

These tests of our formalism and code establish the consistency of our
results with the earlier works mentioned above, and set the stage for
new calculations for arbitrary values of the cosmological constant.
Although our code allows calculations for any dimension, we focus on the
$5-$dimensional case mainly because results already exist for 3, 4 and 6
dimensions.

With $\Lambda =0$, for the tanh initial data, we find a critical exponent 
of $\gamma=0.41$ (Figure 3). This value falls between the 4 and 6 dimensional 
cases as conjectured in \cite{garf_sixd}.

%FIG 3
\begin{figure}[t]
\epsfxsize=130mm
\epsffile{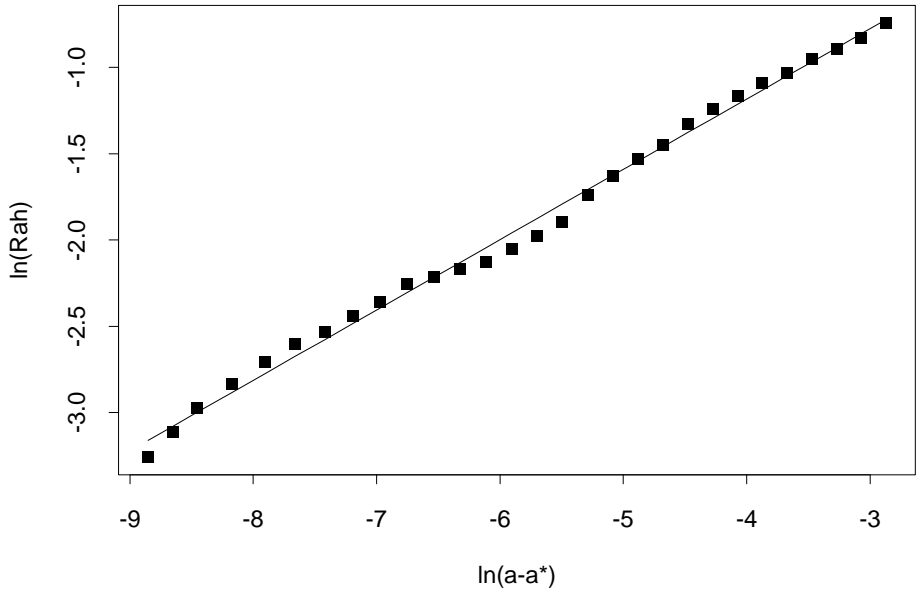}
\caption{\baselineskip = 1.0em Logarithmic plot of the apparent horizon
radius $R_{ah}$ versus
initial scalar field amplitude $(a-a_*)$  in five
spacetime dimensions with zero cosmological constant for tanh initial
data.
$\gamma = 0.41$}
\end{figure}

With  $\Lambda = -1$, and Gaussian initial data,
we find a critical exponent of $\gamma =0.49$ (Figure 4).
All of the graphs show an oscillation about the least squares fit line.
This is a known feature for zero cosmological constant, and is
concomitant with discrete self-similarity of the critical solution. Our
results for negative comological constant also show this feature, which
indicates that the critical solution for this case also has discrete
self-similarity.

We find that in five dimensions the critical exponent appears not to be
universal, at least in the supercritical approach to computing it. The
Gaussian
initial data yielded $\gamma = 0.52$ for  $\Lambda = 0$, in comparison
to
$\gamma = 0.41$ for the tanh initial data (Figure 5).

% FIG 4
\begin{figure}[t]
\epsfxsize=130mm
\epsffile{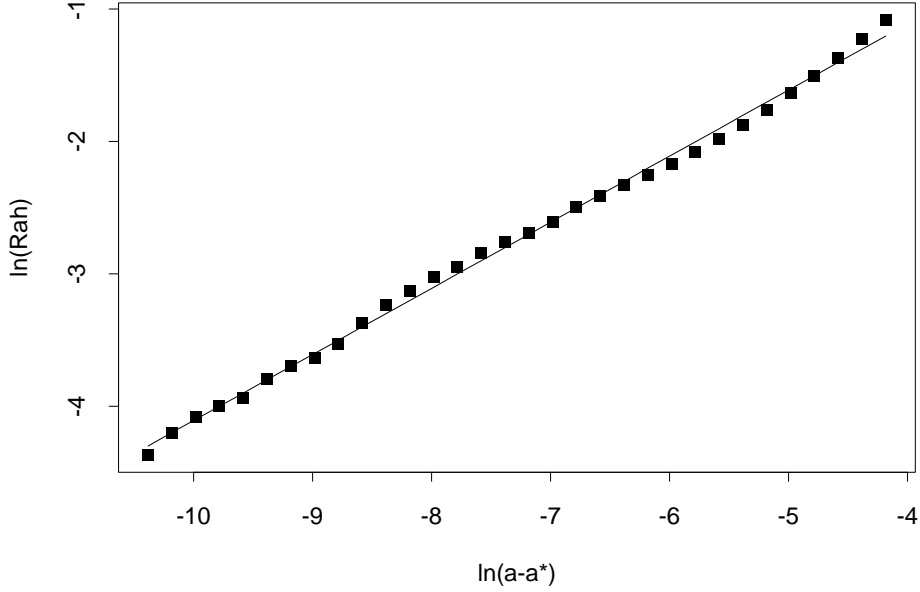}
\caption{\baselineskip = 1.0em Logarithmic plot of  apparent horizon
radius $R_{ah}$ versus
initial scalar field amplitude $(a-a_*)$ in five
spacetime dimensions with $\Lambda = -1$ for Gaussian initial data.
$\gamma = 0.49$}
\end{figure}

%FIG 5
\begin{figure}[t]
\epsfxsize=130mm
\epsffile{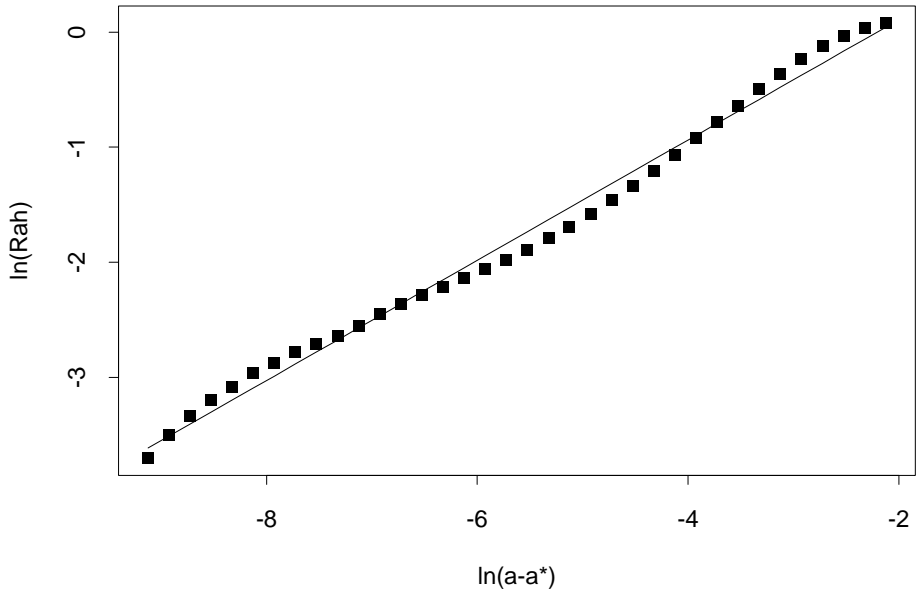}
\caption{\baselineskip = 1.0em Logarithmic plot of  apparent horizon
radius $R_{ah}$ versus
initial scalar field amplitude $(a-a_*)$ in five spacetime dimensions
with
$\Lambda = 0$ for Gaussian initial data. $\gamma = 0.52$}
\end{figure}

The reason for this
is not clear to us and it would be worthwhile calculating the exponents
using the subcritical approach suggested in Ref. \cite{garfR}, where the
Ricci scalar at the origin is calculated near criticality from below. It
is also worth noting that a similar lack of universality is manifested
in the $3-$dimensional AdS case using the supercritical
apparent horizon method of computing $\gamma$ \cite{fransp}.

We also carried out a preliminary investigation of the positive
cosmological constant case in five dimensions. This is an interesting
case because of the presence of a cosmological horizon in addition to
the potential apparent horizon. Figure 6 shows graphs of the scalar
field and apparent horizon functions $h$ and $ah$ in the
left and right columns respectively, as functions of $\phi$, prior to
and at the onset of apparent horizon formation in the two successive
rows. Note the location of the cosmological horizon in the right hand column
near $\phi = 5.8$. We find that the function $\phi(u,v)$ evolved such
that instead of the radial grid contracting
as for the zero and negative $\Lambda$ cases, it expanded as the scalar
field moved towards the origin. This feature is visible in Figure 6: the
range of $\phi$ in the lower graphes has expanded to 8 from 6.  In  fact
the closer is the onset of apparent horizon formation, the larger the range
of the $\phi$ variable (and hence the radial grid). This prevented us from
extracting accurate apparent horizon radii since the interesting features 
became confined to an ever shrinking part of the grid. We hope to study this 
in detail in future work.

%FIG 6
\begin{figure}[t]
\epsfxsize=126mm
\epsffile{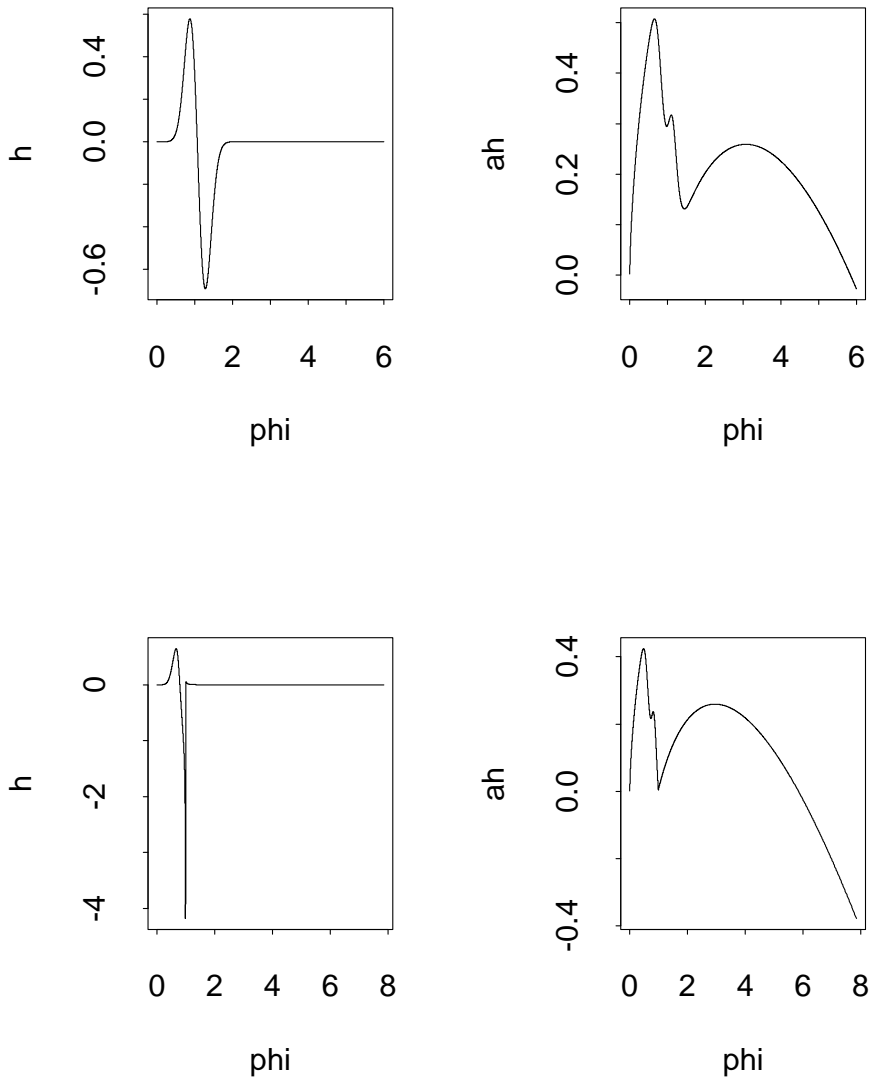}
\caption{\baselineskip = 1.0em Plots of the scalar field $h$ and the
$ah$
function Eqn. (\ref{aheqn}) prior to (top row)  and nearer apparent
horizon formation in five
spacetime dimensions with positive $\Lambda$. Note the expansion of the
$\phi$ grid from 6 to 8 in the bottom two graphs.}
\end{figure}

In summary, we have described a formalism and code for studying
spherically symmetric gravitational collapse of a scalar field
for any $d$ and $\Lambda$, presented new results in five dimensions, 
and given evidence for universality in six dimensions. In future work we
will present results of a systematic analysis of the critical exponent
as a function of both spacetime dimension and cosmological constant.

\section{Acknowledgements}
\par
This work was supported in part by the Natural Sciences and Engineering
Research Council of Canada. G.K. would like to thank J. Gegenberg for
helpful conversations.
  \par

\vspace*{20pt}

\begin{thebibliography}{99}
\bibitem{choptuik} M. Choptuik, Phys. Rev. Lett. {\bf 70}, 9 (1993).
%
\bibitem{gund} C. Gundlach, Living Rev. Rel. 2, 4 (1999).
%
\bibitem{lehn} L. Lehner, Class. Quant. Grav. 18,  R25-R86 (2001).
%
\bibitem{hmn1} V. Husain, E. Martinez, and D. Nunez, Class. Quant. Grav.

{\bf 13}
1183 (1996).
%
\bibitem{hmn2} V. Husain, E. Martinez, and D. Nunez, Phys. Rev. {\bf
D50} 3783 (1994).
%
\bibitem{DG} D. Garfinkle, Phys. Rev. {\bf D51}, 5558 (1995).
%
\bibitem{garf_sixd} D. Garfinkle, C. Cutler and G. Duncan, Phys. Rev.
{\bf D60},
104007 (1999).
%
\bibitem{pc} F. Pretorius and M. Choptuik, Phys. Rev. {\bf D62} 124012
(2000).
%
\bibitem{husain} V. Husain and M. Olivier, Class. Quant. Grav {\bf 18}
L1-L10
(2001).
%
\bibitem{soda} J. Soda and K. Hirata, Phys. Lett. {\bf B387} 271 (1996).

%
\bibitem{frolov} A. Frolov, Class. Quant. Grav. {\bf 16} 407 (1999).
%
\bibitem{dil_grav}J. Gegenberg, D. Louis-Martinez and G. Kunstatter,
Phys. Rev. {\bf D51} 1781 (1995); D. Louis-Martinez and G. Kunstatter,
Phys. Rev. {\bf D52} 3494 (1995).
%
\bibitem{GP} D. Goldwirth and T. Piran, Phys. Rev. {\bf D36}, 3575
(1987).
%
\bibitem{wald} R.M. Wald and V. Iyer, Phys. Rev. {\bf D44}, 3719 (1991).

%
\bibitem{garfR} G. Garfinkle and C. Duncan, Phys. Rev. {\bf D58} (1998)
064024.
%
\bibitem{fransp} F. Pretorius, personal communication.
%
\bibitem{lehn2} L. Lehner, Int. J. Mod. Phys. {\bf D9} 459 (2000). 

\end{thebibliography}
\end{document}